\documentclass{nature}

\usepackage{amsmath,amssymb}
\usepackage{bm}
\usepackage{color}
\usepackage{graphicx}
\usepackage{float}
\usepackage{verbatim}
\usepackage[english]{babel}
\usepackage[colorlinks=true,bookmarks=false,citecolor=blue,urlcolor=blue]{hyperref}
\usepackage{epstopdf}
\usepackage{caption}

\newcommand{\ket}[1]{\left|#1\right>}      
\newcommand{\eps}{\varepsilon}      
\newcommand{\kap}{\varkappa}      

\title{Controlling scattering of light through topological  transitions in all-dielectric metasurfaces}

\author{Maxim A. Gorlach$^{1,2}$, Xiang Ni$^{1,3}$, Daria A. Smirnova$^1$, Dmitry Korobkin$^1$, Alexey P. Slobozhanyuk$^2$, Dmitry Zhirihin$^2$, Pavel A. Belov$^2$, Andrea Al{\`u}$^4$ \& Alexander B. Khanikaev$^{1,2,3}$}

\begin{document}

\maketitle

\begin{affiliations}
\item The City College of the City University of New York, New York 10031, USA

\item ITMO University, Saint Petersburg 197101, Russia

\item Graduate Center of the City University of New York, New York, 10016, USA

\item The University of Texas at Austin, Austin, Texas 78712, USA 
\end{affiliations}

\begin{abstract}
Topological phase transitions in condensed matter systems have shown extremely rich physics, unveiling such exotic states of matter as topological insulators, superconductors and superfluids. Photonic topological systems open a whole new realm of research exhibiting a number of important distinctions from their condensed matter counterparts. 
Photonic modes can couple to the continuum of free space modes which makes it feasible to control and manipulate scattering properties of the photonic structure via topology. At the same time, the direct connection of  scattering and topological properties of the photonic states allows their probing by spectroscopic means via Fano resonances.
Here we demonstrate that the radiative coupling of modes supported by an all-dielectric metasurface can be controlled and tuned under topological phase transitions due to band inversion, correspondingly inducing a distinct switching of the quality factors of the resonances associated with the bands. In addition, we develop a technique to retrieve the topological properties of all-dielectric metasurfaces from the measured far-field scattering characteristics. The collected angle-resolved transmission and reflection spectra allow extracting the momentum-dependent frequencies and lifetimes of the photonic modes. This enables retrieval of the effective photonic Hamiltonian, including the effects of a synthetic gauge field, and topological invariants~-- pseudo-spin Chern numbers. Our results thus open a new avenue to design a new class of metasurfaces with unique scattering characteristics controlled via topological effects. This work also  demonstrates how  topological states of open systems can be explored via far-field measurements.
\end{abstract}

\maketitle

Topological phase transitions in two-dimensional (2D) condensed matter systems have attracted an enormous interest~\cite{Kosterlitz,Bernevig,Lutchyn,CarusottoPRX} crowned by the Nobel Prize in Physics in 2016. With the advent of photonic topological insulators~\cite{Raghu,Wang,Hafezi-11, Khanikaev-13,Hafezi-13,Rechtsman,Ozawa,Slob-NP} the research domain expanded to include topological phase transitions for light~\cite{Lu2014,Lu2016}. However, most of the topological photonic systems considered until today have largely ignored the fact that photonic modes can couple to the continuum of free-space modes, and their exploration has been limited to near-field properties. Only recently topological phenomena in photonic systems associated with leaky states have attracted attention in the context of non-Hermitian Floquet systems~\cite{Zeuner2015,Weimann2016} and non-radiative modes in the continuum~\cite{Hsu2016}. 

While the leakage of photonic modes to the free space continuum can be considered as a challenge, as it makes the system non-Hermitian, it may also offer an alternative route to explore the topology of photonic bands with far-field measurements, provided that the radiative decay of the topological modes is controllable and not destructive. This is especially relevant in the context of metasurfaces --  ultrathin photonic structures enabling a variety of novel optical devices including flat lenses, and wave-plates to control polarization and angular momentum of optical beams~\cite{YuReview,ShalaevScience13,MinovichReview,Bliokh2015,Glybovski2016,Xiao2017}. Endowing metasurfaces with topological properties may open a whole new realm of opportunities in the field of light scattering and wavefront control, as the synthetic gauge field for light have been shown to controllably affect all degrees of freedom associated with electromagnetic radiation. Provided the metasurface is engineered to sustain topological photonic modes above the light line, they may couple to the radiative continuum, giving rise to unique scattering features, such as Fano resonances discriminating light by its polarization, angular momentum, or any other synthetic degree of freedom.

In recent years, it has become a traditional approach to look at the topological properties of photonic systems through the prism of the bulk-boundary correspondence principle~\cite{BernevigBook}, investigating their edge characteristics, including chiral and helical edge states. While this strategy to explore topological properties~\cite{Wang,Hafezi-13,Ma,Slob-16} is quite appealing, due to their distinctive features, backscattering immune propagation in first place, it can divert us from potentially beneficial properties of bulk modes stemming from their topological nature. 
 To explore these ideas, in this Article, we design and fabricate a metasurface that supports radiative photonic modes in the near-infrared spectral range and exhibits a topological transition. As the  parameters of the metasurface are tuned, the coupling of topological modes to the radiation continuum enables the observation of a topological transition, which is accompanied by the inversion of bright and dark modes, referred to as band crossing. Angle-resolved spectroscopy allows the direct extraction of spectral positions, intensity and width of the corresponding peaks in the transmittance spectrum. As demonstrated below, these parameters provide valuable information about the nature of the eigenmodes supported by the structure, enabling the univocal retrieval of topological invariants from far-field measurements.

   \begin{figure}
    \begin{center}
    \includegraphics[width=1\linewidth]{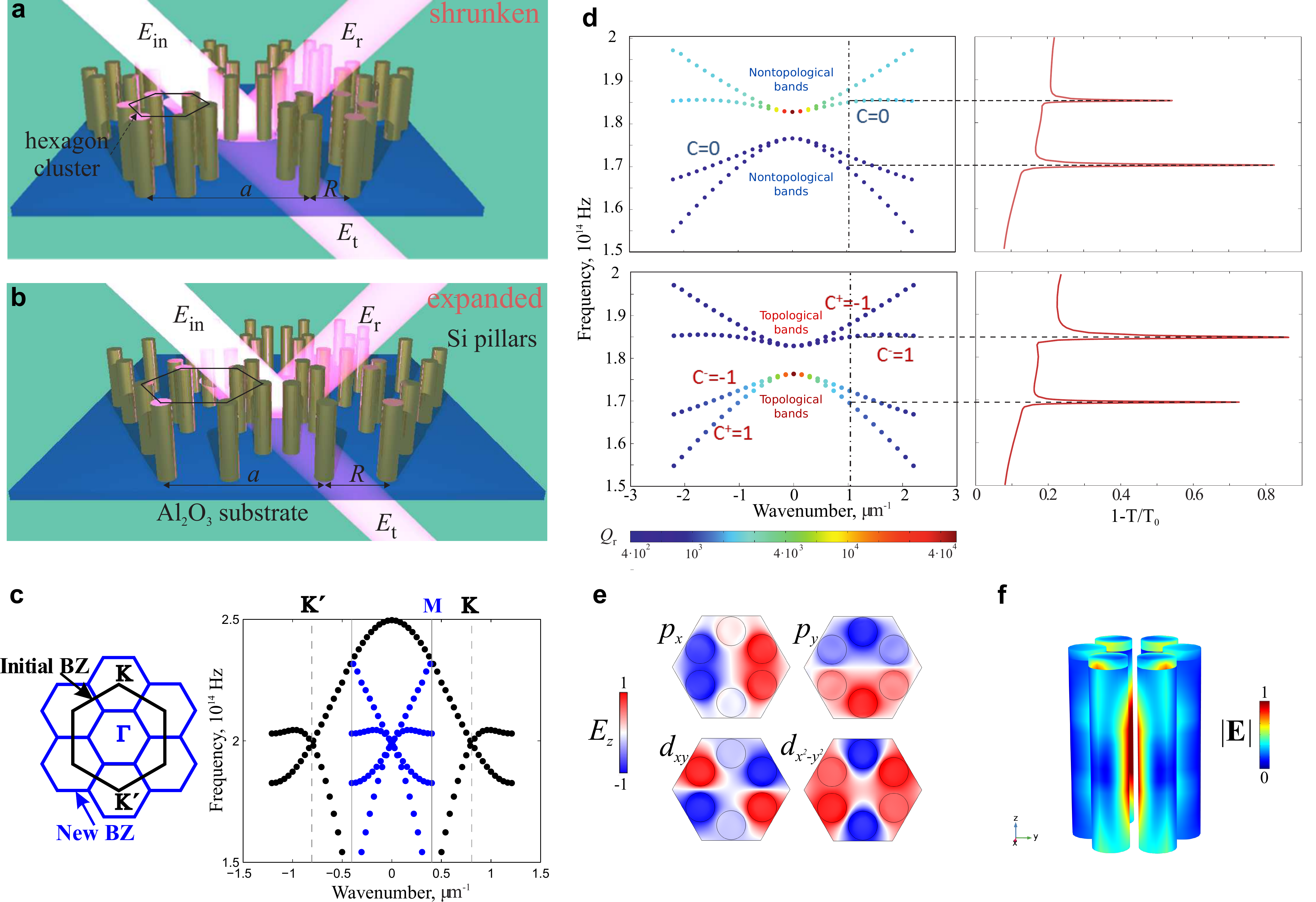}
    \caption{{\bf Design of a photonic metasurface exhibiting topological transition and band inversion above the light line.} (a,b) Geometry of a metasurface composed of a triangular lattice of hexamers of Si pillars on a sapphire ($\text{Al}_2\text{O}_3$) substrate: (a) shrunken structure with $a/R>3$; (b) expanded structure with $a/R<3$. (c) Left: the  Brillouin zone of an unperturbed honeycomb lattice $a/R=3$ (black) and the triangular lattice under study $a/R\neq3$ (blue) obtained by the honeycomb lattice deformation. Right: Shrinking/expanding the hexamer leads to folding of K and K' points to $\Gamma$ point in the new lattice. As the result of such symmetry reduction, valleys (pseudo-spins) mix, which may lead to a topological transition. (d) Demonstration of band inversion. Left column: Complex photonic band structure for the four doublet bands of shrunken $a/R=3.15$ (top), and expanded $a/R=2.85$ (bottom), structures. Color encodes the radiative quality factor of the modes. Right column: Extinction spectrum $1-T/T_0$ of the metasurface normalized to the transmittance $T_0$ of the Al$_2$O$_3$ substrate computed for the fixed angle of incidence $16^{\circ}$. The peaks in far-field spectrum correspond to the eigenmode frequencies for the given tangential wave number. 
(e) Simulated field profiles showing the $E_z$ field component at the top surface of the unit cell for dipolar (top row) and quadrupolar (bottom row) eigenmodes for expanded (topological) structure, (f) Side view of the field distribution in the unit cell under resonant excitation of the dipolar mode at normal incidence by $x$-polarized light.  electric field magnitude is normalized to the maximum value. 	 
 \label{fig:DistortedLattice}}
    \end{center}
    \end{figure}

The metasurface is based on cylindrical Si pillars arranged in hexagon clusters with edge length $R$, placed at the sites of a triangular lattice with period $a$ [Fig.~\ref{fig:DistortedLattice}(a,b)]. 
The topological properties of its infinite 2D analogue possessing $C_{6v}$ symmetry have been the subject of several recent studies~\cite{Wu,Hafezi-16,Dong,Ma-16,Rechtsman17}. For $a/R=3$, this system is a conventional honeycomb lattice with unit cell formed by two cylinders, which exhibits Dirac cones centered at K and $\text{K}^\prime$ points of the Brillouin zone. However, if the lattice symmetry is reduced by clustering six neighboring pillars so that $a/R\not= 3$ (distorted lattice), the size of the unit cell increases leading to the reshaping of the first Brillouin zone [Fig.~\ref{fig:DistortedLattice}(c)]. As a consequence, the Dirac points appear in the vicinity of the $\Gamma$ point due to band folding in the distorted lattice. Additionally, the interaction between the valleys of the former honeycomb lattice caused by such symmetry reduction leads to the opening of photonic bandgaps. This interaction can be viewed as synthetic spin-orbit coupling (gauge field) between pseudo-spins of former valley degrees of freedom. Previous studies suggested that the shrunken structure with $a/R>3$  is topologically trivial, whereas the expanded system with $a/R<3$  is topologically nontrivial~\cite{Wu} (see also Sup.~Mat.~Sec.~I), which, as shown below, remains true for the case of the open system. The important distinction however is that in the case of the open metasurface the folded modes appear within the light cone and therefore are radiatively coupled to the continuum of free space, which enables their far field characterization~\cite{Regan16}.

To perform spectroscopy measurements of metasurfaces with two topologically distinct geometries, we fabricated two sets of samples [Fig.~\ref{fig:DistortedLattice}(a,b)] with the same lattice period $a=750$~nm, radius of silicon pillars $r=75$~nm and height of the pillars $h=1.0~\mu\text{m}$. The sizes of the clusters $R$ shown in Fig.~\ref{fig:DistortedLattice}(a,b) were chosen to be $a/R=3.15$ and $a/R=2.85$ for shrunken and expanded structures, respectively. 

{\bf Effective Hamiltonian.} 
While the topological properties of the perturbed honeycomb lattice have been previously explored under the tight binding approximation, here we take a different approach, fully considering the electromagnetic nature of the system. Correspondingly, the effective Hamiltonian near the $\Gamma$ point of the Brillouin zone is obtained directly from Maxwell's equations by the plane wave expansion method~\cite{Khanikaev-13,Slob-NP} (Sup.~Mat. Sec.~II). To describe the properties of photonic bands of bulk modes in the 2D photonic crystal in the vicinity of the $\Gamma$ point, we construct an effective $4\times 4$ Hamiltonian. 
 Focusing on TM-polarized modes with $E_z$ component of the electric field directed along the Si rods' axis, we find the results in compliance with the tight binding model (Sup.~Mat. Sec.~I and Refs.~\cite{Wu,Hafezi-16}). The derived effective $4\times 4$ Hamiltonian has the following form~\cite{Wu}:
\begin{equation}\label{EffHam1}
\hat{H}=
\begin{pmatrix}
\hat{H}_{-} & \hat{K}\\
\hat{K}^{\dag} & \hat{H}_{+}
\end{pmatrix}
\:,
\end{equation}
where the $2\times 2$ matrices $\hat{H}_{\pm}$ and $\hat{K}$ read
\begin{equation}\label{Hpm}
\hat{H}_{\pm}=
\begin{pmatrix}
\mu(k) & v\,(\mp k_x-i\,k_y)\\
v\,(\mp k_x+i\,k_y) & -\mu(k)
\end{pmatrix}
\:,
\end{equation}
\begin{equation}\label{Coupling}
\hat{K}=
\begin{pmatrix}
\alpha\,(k_x+ik_y)^2 & 0\\
0 & -\alpha\,(k_x-i\,k_y)^2\\
\end{pmatrix}
\:,
\end{equation}
$\mu(k)=\mu+\beta\,k^2$, $\mu$ and $\beta$ are the mass term and band parabolicity, respectively.
This form of the Hamiltonian corresponds to the basis choice
\begin{equation}\label{Psi}
\ket{\psi}=\left(\ket{p_-},\ket{d_-},\ket{p_+},\ket{d_+}\right)^T\:,
\end{equation}
where, $\ket{p_{\pm}}=p_x\pm i p_y$ denote circularly polarized dipolar modes and $\ket{d_{\pm}}=  d_{x^2-y^2}\pm i d_{xy}$ stand for the "circularly polarized" quadrupolar modes, which both originate from TM modes supported by an isolated meta-molecule (the hexamer of 6 dielectric rods). 

The four eigenstates of the Hamiltonian Eq.~\eqref{EffHam1} 
exhibit a pairwise degeneracy at $\Gamma$ point $k=0$ (Sup.~Mat.~Sec.~I, II), with eigenvalues equal to $\mu$ (dipolar modes) and $-\mu$ (quadrupolar modes). Topological transition from shrunk to expanded design is characterized by a band inversion for dipolar and quadrupolar eigenstates, accompanied by an inversion of the sign of the mass term $\mu$. The degeneracy is removed for nonzero values of $k$ due to the term $\propto \alpha$, which describes the coupling of left- and right-handed  circularly polarized modes. However, for the topological invariant calculation, the term proportional to $\alpha$ is inessential and it can be dropped~\cite{Wu}. As a result of this approximation, the Hamiltonian splits into two independent $2\times 2$ blocks, and the effective pseudo-spin can be introduced. A straightforward calculation~\cite{Shen} of spin Chern number (Sup.~Mat.~I) yields
\begin{equation}\label{SpinChern}
C=\frac{1}{2}\,\left[\text{sgn}\,\mu-\text{sgn}\,\beta\right]\:.
\end{equation}
Thus, to evaluate the topological invariant it is sufficient to extract the mass term $\mu$ and the parabolicity of the bands $\beta$ from the experimentally measured spectra.

{\bf Retrieval of topological order through light scattering by the metasurface.} 
The modes of interest are located above the light cone and, for the case of a metasurface of a finite thickness within an open background, they couple to the radiation continuum~\cite{Regan16}. As a result, the modes have a finite lifetime, and the corresponding eigenfrequencies become complex valued. Thus, the Si pillars constituting the metasurface effectively act as laterally coupled cavities, and the leaky modes of the structure can be excited by the polarization currents induced at the interface of the metasurface by the incident fields. 
However, taking into consideration the symmetry of the modes, normally incident light can only couple to the dipolar modes, whereas coupling to the quadrupolar modes is suppressed because of the symmetry mismatch with the incident field. Nonetheless, the quadrupolar modes can be excited indirectly through coupling to dipolar modes at oblique incidence, because of the hybridization of dipolar and quadrupolar modes away from the $\Gamma$ point. This mechanism is analogous to the extrinsic coupling to dark modes in Fano-resonant systems caused by symmetry reduction due to the finite value of the in-plane wave vector components. 
Of special interest are the effects of topological transition on the radiative coupling of the modes to the continuum, which should follow the band inversion described above, giving rise to switching of the bright (dipolar) and dark (quadrupolar) modes thus enabling control over the far-field scattering properties of the metasurface.

To support our claims, the complex dispersion diagrams for 
both the shrunken and expanded metasurfaces were calculated in Comsol Multiphysics with radiative decay fully considered. The results are shown in Fig.~\ref{fig:DistortedLattice}(d), left. Indeed, the dipolar and quadrupolar modes at the $\Gamma$ point appear to be bright and dark, respectively, as reflected by the radiative quality factors. However, even at oblique incidence the respective modes may have radiative quality factors that differ by orders of magnitude. The numerically calculated reflectance spectra at oblique incidence are shown in Fig.~\ref{fig:DistortedLattice}(d), right, and confirm excitation of the two eigenmodes, which manifest as two peaks. The predominantly dipolar and quadrupolar modes can be easily discriminated by their distinct bandwidth and the amplitude of the corresponding peaks.

To formalize our description in the context of the topological properties of leaky modes, we develop a unified approach utilizing the temporal coupled mode theory (CMT) along with the electromagnetic effective Hamiltonian description. We write the CMT equations for the amplitudes of the right- and left-handed circularly polarized leaky modes coupled to the external source in the following block-diagonal form:
\begin{equation}\label{Coupled}
-i\eps\,\ket{\psi_{\pm}}=-i\,\hat{H}_{\pm}\,\ket{\psi_{\pm}}+\kap\,
\begin{pmatrix}
E_{\rm{in}}\\ 0
\end{pmatrix}
-
\begin{pmatrix}
\gamma_0+\gamma_r& 0\\
0 & \gamma_0
\end{pmatrix}
\,\ket{\psi_{\pm}}
\end{equation}
where $\ket{\psi_{\pm}}=\left(\ket{p_{\pm}},\ket{d_{\pm}}\right)^T$ is the ``wavefunction'' composed of $p$ (dipole) and $d$ (quadrupole) modes of the system. 
The first term of Eq.~\eqref{Coupled} describes the evolution of the coupled modes in a closed system, $\kap$ describes the coupling strength of the system to the external field $E_{\rm{in}} = E_0/\sqrt{2}$, $\gamma_0$ represents non-radiative losses in the structure, and $\gamma_{\rm{r}}=\kap^2/2$ describes the radiative losses. The use of this form of dynamic equation can be also justified by applying electromagnetic perturbation theory for an open system as shown in Sup.~Mat.~IV. 

Once Eq.~\eqref{Coupled} is solved,
one can obtain expressions for transmission (and reflection) coefficients:
\begin{equation}\label{Transmitted-2}
t_{\pm}=t_0\,\left[1-\kap\,\psi_{\pm}(p)/E_{\rm{in}}\right]\:,
\end{equation} 
where $\psi_{\pm}(p)$ is the first (dipolar) component of the two-component wave function $\ket{\psi_{\pm}}$.
Since the results are the same for left- and right-handed circular polarizations, from now on we omit the $\pm$ subscript. Plugging the Hamiltonian into Eq.~\eqref{Transmitted-2}, we derive the expression for the normalized reflectivity
\begin{equation}\label{Transmitted-3}
\tilde{R}\equiv 1-\frac{|t|^2}{|t_0|^2}=2\,\gamma_0\,\kap^2\,\frac{(\eps+\mu(k))^2+v^2\,k^2+\gamma_0^2}{\left[\mu(k)^2-\eps^2+v^2\,k^2+\gamma_0\,\left(\gamma_0+\kap^2/2\right)\right]^2+\left[2\gamma_0\,\eps+\kap^2/2\,(\mu(k)+\eps)\right]^2}\:.
\end{equation}
Equation~\eqref{Transmitted-3} suggests in particular that two peaks in the $\tilde{R}$ spectrum can be observed due to excitation of dipolar and quadrupolar eigenmodes. Thus, the frequency of the most pronounced peak at incidence close to normal is around  $f_0+\mu$, whereas the frequency of the less intense peak is found around $f_0-\mu$, where $f_0$ is the center-of-bandgap frequency.  

Overall, there are six parameters in the effective model that control the entire dispersion and topological properties of the far-field response: $\mu$, $\gamma_{\rm{r}}=\kap^2/2$, $\gamma_0$, $v$ and $\beta$. The numerical values of parameters comprising the effective Hamiltonian can be obtained by fitting the experimental data. Next, by using the developed technique, we analyze the experimental data for both shrunken ($a/R=3.15$) and expanded ($a/R=2.85$) structures and extract their topological characteristics.

\section*{Results}

\begin{figure}
    \begin{center}
    \includegraphics[width=0.9\linewidth]{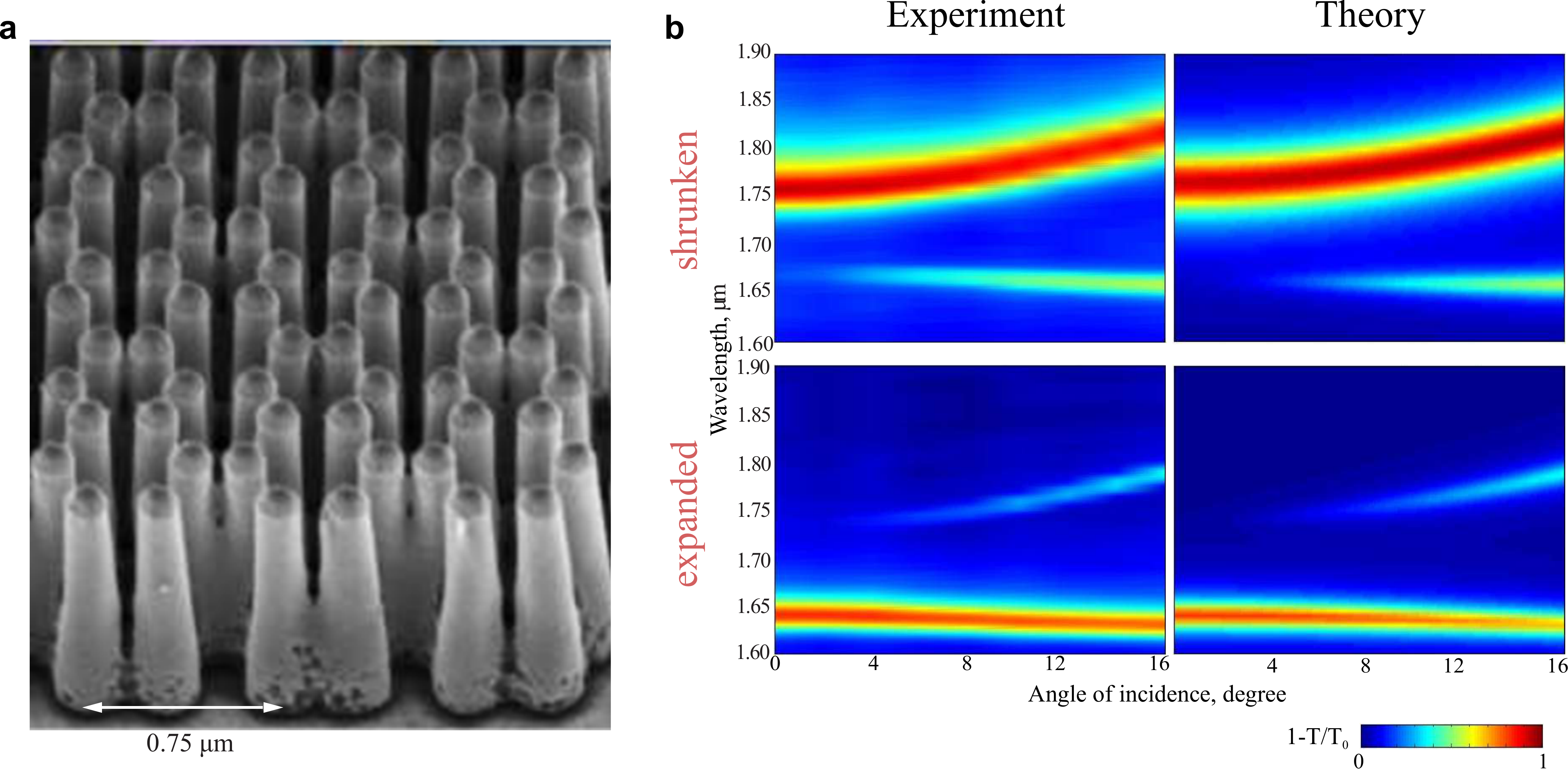}
    \caption{ {\bf Measured reflectance spectra for the fabricated dielectric metasurface.} (a) Scanning electron microscopy image of the fabricated expanded structure with lattice period $a=0.75~\mu$m, $a/R=2.85$, pillar height $h=1.0~\mu$m and pillar radius $r=75$~nm. 
		(b) Experimental vs numerical spectra for the fabricated shrunken (top) and expanded (bottom) samples with $a/R=3.15$ and $a/R=2.85$, respectively. Color encodes the magnitude of reflectance for $p$-polarized incident light. \label{fig:ColorMaps}}
    \end{center}
    \end{figure}

In our proof-of-concept experiment, the structure was illuminated using the source Ocean Optics HL-2000-LL. The transmission coefficient was measured for both structures by the spectrometer Ocean Optics NIR Quest NQ 512-2.2 in the range of wavelengths from $896$ to $2142$~nm corresponding to the frequency range $\left(1.40\div 3.35 \right)\cdot 10^{14}$~Hz. The measured transmittance was normalized to the one of a sapphire substrate.
Figure~\ref{fig:ColorMaps} shows the color map of the quantity $\tilde{R}=1-T/T_0$ as a function of wavelength and incidence angle for both shrunken and expanded structures.

In the wavelength range from $1.6~\mu\text{m}$ to $1.9~\mu\text{m}$, two peaks are clearly observed. The spectroscopy results for the whole studied spectral range are provided in Sup.~Mat.~III. In agreement with our numerical calculations, the experimental data shows clearly that the most intense and broad peak for the shrunken structure is the low-frequency one, whereas for the expanded structure the high-frequency peak is more pronounced. Thus, the shrunken structure is clearly characterized by a negative effective mass $\mu$, while the expanded structure is described by a positive effective mass $\mu$.

\begin{figure}
\begin{center}
\includegraphics[width=0.7\linewidth]{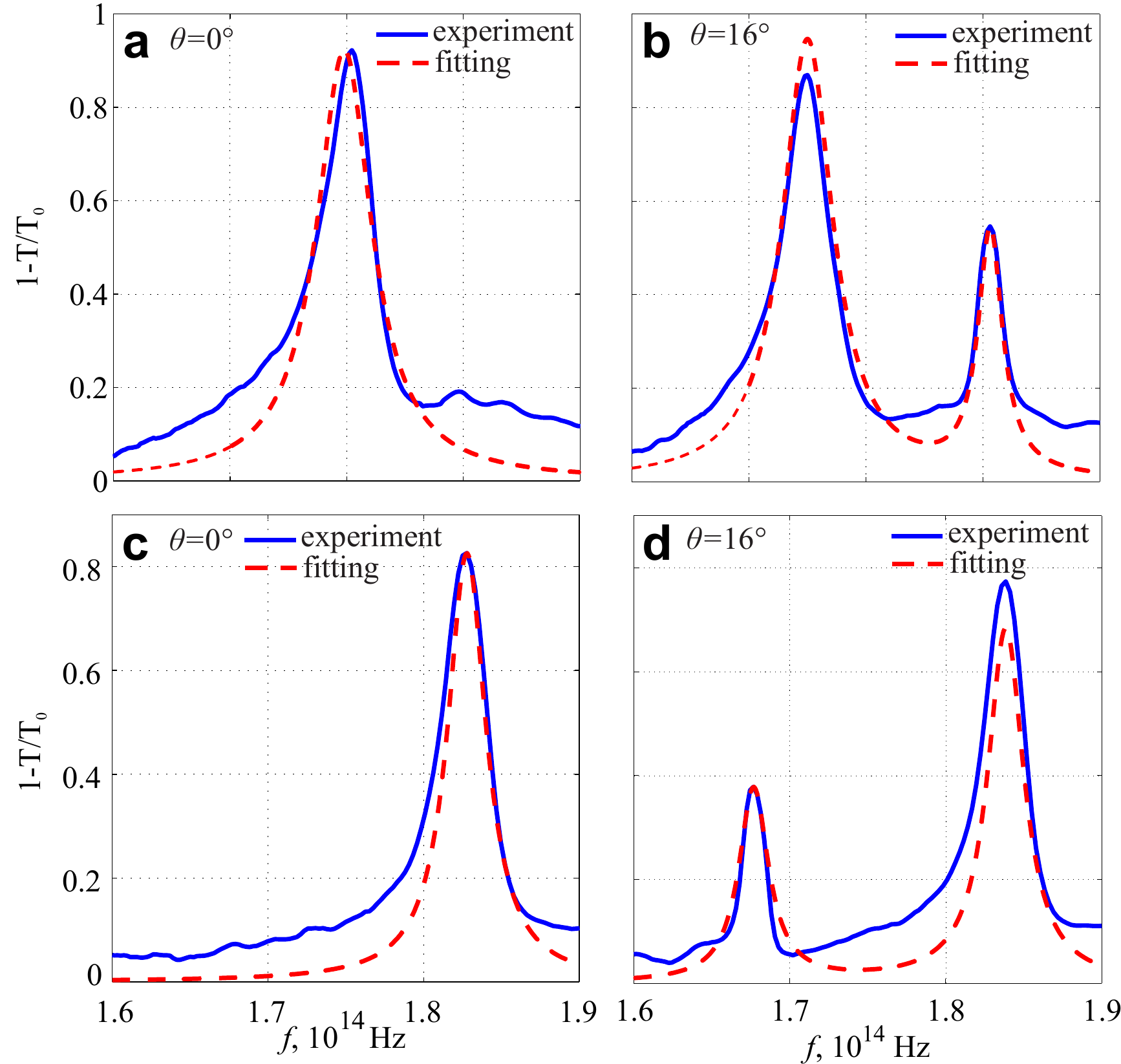}
\caption{Fitting of the experimental spectra ($1-T/T_0$, where $T_0$ is the substrate transmittance) by the analytical model Eq.~\eqref{Transmitted-3} for (a,b) shrunken and (c,d) expanded structures with $a/R=3.15$ and $a/R=2.85$, respectively, at two different angles (a,c) $\theta=0$ and (b,d) $\theta=16^{\circ}$. \label{ris:Fitting}}
\end{center}
\end{figure}

Then we applied the above model to fit the measured data with Eqs.~\eqref{Coupled}-\eqref{Transmitted-3}, as detailed in the Sup.~Mat.~III. The results of the simplest fitting algorithm for two particular angles of incidence, 0 and 16~deg,  are illustrated in Fig.~\ref{ris:Fitting} for the cases of both shrunken and expanded structures. The extracted parameters of the effective Hamiltonian are listed in Table~1. Importantly, these numbers appear to be nearly independent of the fitting algorithm (see Sup.~Mat.~III). From the measured reflectance spectra, we recover again that the shrunken structure is topologically trivial, since the effective mass $\mu$ and parabolicity parameter $\beta$ have the same sign, yielding zero spin Chern number. On the contrary, the expanded structure appears to be topologically nontrivial, due to the opposite signs of $\mu$ and $\beta$, yielding a spin Chern number of 1. These results confirm that it is possible to retrieve the topological nature of a metasurface, and the band inversion properties, with far-field measurements in a robust and direct way.

\begin{center}
\captionof{table}{}
\begin{tabular}[b]{|p{10em}|p{3em}|p{3em}|p{4em}|p{3em}|p{3em}|p{6em}|}
\hline
Structure & $\mu$, THz & $\beta$, $\text{m}^2/\text{s}$ & $v$, $10^6$~m/s & $\gamma_{\rm{r}}$, THz & $\gamma_0$, THz & $C$ (Spin Chern number) \\
\hline
Shrunken, $a/R=3.15$ & -5.22 & -1.00 & 4.76 & 1.83 & 1.03 & 0 \\
\hline
Expanded, $a/R=2.85$ & 5.22 & -1.96 & 7.31 & 0.437 & 1.06 & 1 \\
\hline
\end{tabular}
\label{tab:params}
\end{center}

\section*{Conclusion}

In this paper, we introduced the concept of topological metasurfaces, and demonstrated that their scattering characteristics, radiative quality factors of the modes in particular, can be controlled by synthetic gauge fields. The developed formalism allows us to relate topological and far-field scattering properties enabling the extraction of the effective Hamiltonian and topological invariant (spin Chern number) of the metasurface from the measured transmission spectra. Coupling of photonic modes of the structure to the free-space modes allows  probing directly the topological phase transitions in the open photonic system. We believe that our results open a new avenue for designing metasurfaces with desirable scattering characteristics, topologically robust to continuous modifications and disorder, by engineering synthetic degrees of freedom of light and gauge fields acting on them.

\section*{Methods}

To derive the effective Hamiltonian and topological properties of the photonic crystal, we use tight binding approach (Sup.~Mat.~I) as well as plane wave expansion method (Sup.~Mat.~II). To calculate the reflection and transmission properties of the structure, we employ coupled mode theory (Sup.~Mat.~III). Perturbative analysis of the radiative losses for the metasurface is performed by the guided mode expansion method (Sup.~Mat.~IV). A scheme of the fabrication procedure is outlined in Sup.~Mat.~V. 


\begin{addendum}
 \item The work was supported by the National Science Foundation grants CMMI-1537294 and EFRI-1641069. Research was partly carried out at the Center for Functional Nanomaterials, Brookhaven National Laboratory, which is supported by the US Department of Energy, Office of Basic Energy Sciences, under contract no. DE-SC0012704. The theoretical models were developed with the support of the Russian Science Foundation (Grant No.~16-19-10538).
 \item[Competing Interests] The authors declare that they have no
competing financial interests.
 \item[Correspondence] Correspondence and requests for materials
should be addressed to \\ A.B.K. (email: akhanikaev@ccny.cuny.edu) and A.A. (email: alu@mail.utexas.edu).
\end{addendum}

\bibliographystyle{naturemag}
\bibliography{TopologicalLib3}

\begin{thebibliography}{10}
\expandafter\ifx\csname url\endcsname\relax
  \def\url#1{\texttt{#1}}\fi
\expandafter\ifx\csname urlprefix\endcsname\relax\def\urlprefix{URL }\fi
\providecommand{\bibinfo}[2]{#2}
\providecommand{\eprint}[2][]{\url{#2}}

\bibitem{Kosterlitz}
\bibinfo{author}{Kosterlitz, J.~M.} \& \bibinfo{author}{Thouless, D.~J.}
\newblock \bibinfo{title}{Ordering, metastability and phase transitions in
  two-dimensional systems}.
\newblock \emph{\bibinfo{journal}{Journal of Physics C}}
  \textbf{\bibinfo{volume}{6}}, \bibinfo{pages}{1181--1203}
\newblock  (\bibinfo{year}{1973}).

\bibitem{Bernevig}
\bibinfo{author}{Bernevig, B.~A.}, \bibinfo{author}{Hughes, T.~L.} \&
  \bibinfo{author}{Zhang, S.-C.}
\newblock \bibinfo{title}{Quantum spin {H}all effect and topological phase
  transition in {HgTe} quantum wells}.
\newblock \emph{\bibinfo{journal}{Science}} \textbf{\bibinfo{volume}{314}},
  \bibinfo{pages}{1757--1761}
\newblock  (\bibinfo{year}{2006}).

\bibitem{Lutchyn}
\bibinfo{author}{Lutchyn, R.~M.}, \bibinfo{author}{Sau, J.~D.} \&
  \bibinfo{author}{Sarma, S.~D.}
\newblock \bibinfo{title}{Majorana fermions and a topological phase transition
  in semiconductor-superconductor heterostructures}.
\newblock \emph{\bibinfo{journal}{Physical Review Letters}}
  \textbf{\bibinfo{volume}{105}}, \bibinfo{pages}{077001}
\newblock  (\bibinfo{year}{2010}).

\bibitem{CarusottoPRX}
\bibinfo{author}{Dagvadorj, G.} \emph{et~al.}
\newblock \bibinfo{title}{Nonequilibrium phase transition in a two-dimensional
  driven open quantum system}.
\newblock \emph{\bibinfo{journal}{Phys. Rev. X}} \textbf{\bibinfo{volume}{5}},
  \bibinfo{pages}{041028}
\newblock  (\bibinfo{year}{2015}).

\bibitem{Raghu}
\bibinfo{author}{Raghu, S.} \& \bibinfo{author}{Haldane, F. D.~M.}
\newblock \bibinfo{title}{{Analogs of quantum-Hall-effect edge states in
  photonic crystals}}.
\newblock \emph{\bibinfo{journal}{Physical Review A}}
  \textbf{\bibinfo{volume}{78}}, \bibinfo{pages}{033834}
\newblock  (\bibinfo{year}{2008}).

\bibitem{Wang}
\bibinfo{author}{Wang, Z.}, \bibinfo{author}{Chong, Y.},
  \bibinfo{author}{Joannopoulos, J.~D.} \& \bibinfo{author}{Solja{\v c}i{\'c},
  M.}
\newblock \bibinfo{title}{Observation of unidirectional backscattering-immune
  topological electromagnetic states}.
\newblock \emph{\bibinfo{journal}{Nature}} \textbf{\bibinfo{volume}{461}},
  \bibinfo{pages}{772--775}
\newblock  (\bibinfo{year}{2009}).

\bibitem{Hafezi-11}
\bibinfo{author}{Hafezi, M.}, \bibinfo{author}{Demler, E.~A.},
  \bibinfo{author}{Lukin, M.~D.} \& \bibinfo{author}{Taylor, J.~M.}
\newblock \bibinfo{title}{Robust optical delay lines with topological
  protection}.
\newblock \emph{\bibinfo{journal}{Nature Physics}}
  \textbf{\bibinfo{volume}{7}}, \bibinfo{pages}{907--912}
\newblock  (\bibinfo{year}{2011}).

\bibitem{Khanikaev-13}
\bibinfo{author}{Khanikaev, A.~B.} \emph{et~al.}
\newblock \bibinfo{title}{Photonic topological insulators}.
\newblock \emph{\bibinfo{journal}{Nature Materials}}
  \textbf{\bibinfo{volume}{12}}, \bibinfo{pages}{233}
\newblock  (\bibinfo{year}{2013}).

\bibitem{Hafezi-13}
\bibinfo{author}{Hafezi, M.}, \bibinfo{author}{Mittal, S.},
  \bibinfo{author}{Fan, J.}, \bibinfo{author}{Migdall, A.} \&
  \bibinfo{author}{Taylor, J.~M.}
\newblock \bibinfo{title}{Imaging topological edge states in silicon
  photonics}.
\newblock \emph{\bibinfo{journal}{Nature Photonics}}
  \textbf{\bibinfo{volume}{7}}, \bibinfo{pages}{1001--1005}
\newblock  (\bibinfo{year}{2013}).

\bibitem{Rechtsman}
\bibinfo{author}{Rechtsman, M.~C.} \emph{et~al.}
\newblock \bibinfo{title}{{Photonic Floquet topological insulators}}.
\newblock \emph{\bibinfo{journal}{Nature}} \textbf{\bibinfo{volume}{496}},
  \bibinfo{pages}{196--200}
\newblock  (\bibinfo{year}{2013}).

\bibitem{Ozawa}
\bibinfo{author}{Ozawa, T.}, \bibinfo{author}{Price, H.~M.},
  \bibinfo{author}{Goldman, N.}, \bibinfo{author}{Zilberberg, O.} \&
  \bibinfo{author}{Carusotto, I.}
\newblock \bibinfo{title}{{Synthetic dimensions in integrated photonics: From
  optical isolation to four-dimensional quantum Hall physics}}.
\newblock \emph{\bibinfo{journal}{Phys. Rev. A}} \textbf{\bibinfo{volume}{93}},
  \bibinfo{pages}{043827}
\newblock  (\bibinfo{year}{2016}).

\bibitem{Slob-NP}
\bibinfo{author}{Slobozhanyuk, A.} \emph{et~al.}
\newblock \bibinfo{title}{Three-dimensional all-dielectric photonic topological
  insulator}.
\newblock \emph{\bibinfo{journal}{Nature Photonics}}
  \textbf{\bibinfo{volume}{11}}, \bibinfo{pages}{130--136}
\newblock  (\bibinfo{year}{2016}).

\bibitem{Lu2014}
\bibinfo{author}{{Lu}, L.}, \bibinfo{author}{{Joannopoulos}, J.~D.} \&
  \bibinfo{author}{{Solja{\v c}i{\'c}}, M.}
\newblock \bibinfo{title}{{Topological photonics}}.
\newblock \emph{\bibinfo{journal}{Nature Photonics}}
  \textbf{\bibinfo{volume}{8}}, \bibinfo{pages}{821--829}
\newblock  (\bibinfo{year}{2014}).

\bibitem{Lu2016}
\bibinfo{author}{Lu, L.}, \bibinfo{author}{Joannopoulos, J.~D.} \&
  \bibinfo{author}{Solja{\v c}i{\'c}, M.}
\newblock \bibinfo{title}{Topological states in photonic systems}.
\newblock \emph{\bibinfo{journal}{Nature Physics}}
  \textbf{\bibinfo{volume}{12}}, \bibinfo{pages}{626}
\newblock  (\bibinfo{year}{2016}).

\bibitem{Zeuner2015}
\bibinfo{author}{Zeuner, J.~M.} \emph{et~al.}
\newblock \bibinfo{title}{{Observation of a Topological Transition in the Bulk
  of a Non-Hermitian System}}.
\newblock \emph{\bibinfo{journal}{Phys. Rev. Lett.}}
  \textbf{\bibinfo{volume}{115}}, \bibinfo{pages}{040402}
\newblock  (\bibinfo{year}{2015}).

\bibitem{Weimann2016}
\bibinfo{author}{Weimann, S.} \emph{et~al.}
\newblock \bibinfo{title}{Topologically protected bound states in photonic
  parity{\textendash}time-symmetric crystals}.
\newblock \emph{\bibinfo{journal}{Nature Materials}}
  \textbf{\bibinfo{volume}{16}}, \bibinfo{pages}{433--438}
\newblock  (\bibinfo{year}{2016}).

\bibitem{Hsu2016}
\bibinfo{author}{Hsu, C.~W.}, \bibinfo{author}{Zhen, B.},
  \bibinfo{author}{Stone, A.~D.}, \bibinfo{author}{Joannopoulos, J.~D.} \&
  \bibinfo{author}{Solja{\v{c}}i{\'{c}}, M.}
\newblock \bibinfo{title}{Bound states in the continuum}.
\newblock \emph{\bibinfo{journal}{Nature Reviews Materials}}
  \textbf{\bibinfo{volume}{1}}, \bibinfo{pages}{16048}
\newblock  (\bibinfo{year}{2016}).

\bibitem{YuReview}
\bibinfo{author}{Yu, N.} \& \bibinfo{author}{Capasso, F.}
\newblock \bibinfo{title}{Flat optics with designer metasurfaces}.
\newblock \emph{\bibinfo{journal}{Nat. Mater.}} \textbf{\bibinfo{volume}{13}},
  \bibinfo{pages}{139–150}
\newblock  (\bibinfo{year}{2014}).

\bibitem{ShalaevScience13}
\bibinfo{title}{Planar photonics with metasurfaces}.
\newblock \emph{\bibinfo{journal}{Science}} \textbf{\bibinfo{volume}{339}},
  \bibinfo{pages}{1232009}
\newblock  (\bibinfo{year}{2013}).

\bibitem{MinovichReview}
\bibinfo{author}{Minovich, A.~E.} \emph{et~al.}
\newblock \bibinfo{title}{Functional and nonlinear optical metasurfaces}.
\newblock \emph{\bibinfo{journal}{Laser $\&$ Photon. Rev.}}
  \textbf{\bibinfo{volume}{9}}, \bibinfo{pages}{195}
\newblock  (\bibinfo{year}{2015}).

\bibitem{Bliokh2015}
\bibinfo{author}{Bliokh, K.~Y.},
  \bibinfo{author}{Rodr{\'{\i}}guez-Fortu{\~{n}}o, F.~J.},
  \bibinfo{author}{Nori, F.} \& \bibinfo{author}{Zayats, A.~V.}
\newblock \bibinfo{title}{Spin{\textendash}orbit interactions of light}.
\newblock \emph{\bibinfo{journal}{Nature Photonics}}
  \textbf{\bibinfo{volume}{9}}, \bibinfo{pages}{796--808}
\newblock  (\bibinfo{year}{2015}).

\bibitem{Glybovski2016}
\bibinfo{author}{Glybovski, S.~B.}, \bibinfo{author}{Tretyakov, S.~A.},
  \bibinfo{author}{Belov, P.~A.}, \bibinfo{author}{Kivshar, Y.~S.} \&
  \bibinfo{author}{Simovski, C.~R.}
\newblock \bibinfo{title}{Metasurfaces: From microwaves to visible}.
\newblock \emph{\bibinfo{journal}{Physics Reports}}
  \textbf{\bibinfo{volume}{634}}, \bibinfo{pages}{1--72}
\newblock  (\bibinfo{year}{2016}).

\bibitem{Xiao2017}
\bibinfo{author}{Xiao, S.} \emph{et~al.}
\newblock \bibinfo{title}{Spin-dependent optics with metasurfaces}.
\newblock \emph{\bibinfo{journal}{Nanophotonics}} \textbf{\bibinfo{volume}{6}}
\newblock  (\bibinfo{year}{2017}).

\bibitem{BernevigBook}
\bibinfo{author}{Bernevig, A.~B.} \& \bibinfo{author}{Hughes, T.~L.}
\newblock \emph{\bibinfo{title}{Topological insulators and topological
  superconductors}}
\newblock  (\bibinfo{publisher}{Princeton University Press},
  \bibinfo{year}{2013}).

\bibitem{Ma}
\bibinfo{author}{Ma, T.}, \bibinfo{author}{Khanikaev, A.~B.},
  \bibinfo{author}{Mousavi, S.~H.} \& \bibinfo{author}{Shvets, G.}
\newblock \bibinfo{title}{Guiding electromagnetic waves around sharp corners:
  Topologically protected photonic transport in metawaveguides}.
\newblock \emph{\bibinfo{journal}{Physical Review Letters}}
  \textbf{\bibinfo{volume}{114}}, \bibinfo{pages}{127401}
\newblock  (\bibinfo{year}{2015}).

\bibitem{Slob-16}
\bibinfo{author}{Slobozhanyuk, A.~P.} \emph{et~al.}
\newblock \bibinfo{title}{Experimental demonstration of topological effects in
  bianisotropic metamaterials}.
\newblock \emph{\bibinfo{journal}{Scientific Reports}}
  \textbf{\bibinfo{volume}{6}}, \bibinfo{pages}{22270}
\newblock  (\bibinfo{year}{2016}).

\bibitem{Wu}
\bibinfo{author}{Wu, L.-H.} \& \bibinfo{author}{Hu, X.}
\newblock \bibinfo{title}{{Scheme for Achieving a Topological Photonic Crystal
  by Using Dielectric Material}}.
\newblock \emph{\bibinfo{journal}{Physical Review Letters}}
  \textbf{\bibinfo{volume}{114}}, \bibinfo{pages}{23901}
\newblock  (\bibinfo{year}{2015}).

\bibitem{Hafezi-16}
\bibinfo{author}{Barik, S.}, \bibinfo{author}{Miyake, H.},
  \bibinfo{author}{{DeGottardi}, W.}, \bibinfo{author}{Waks, E.} \&
  \bibinfo{author}{Hafezi, M.}
\newblock \bibinfo{title}{Two-dimensionally confined topological edge states in
  photonic crystals}.
\newblock \emph{\bibinfo{journal}{New Journal of Physics}}
  \textbf{\bibinfo{volume}{18}}, \bibinfo{pages}{113013}
\newblock  (\bibinfo{year}{2016}).

\bibitem{Dong}
\bibinfo{author}{Dong, J.-W.}, \bibinfo{author}{Chen, X.-D.},
  \bibinfo{author}{Zhu, H.}, \bibinfo{author}{Wang, Y.} \&
  \bibinfo{author}{Zhang, X.}
\newblock \bibinfo{title}{Valley photonic crystals for control of spin and
  topology}.
\newblock \emph{\bibinfo{journal}{Nature Materials}}
  \textbf{\bibinfo{volume}{16}}, \bibinfo{pages}{298--302}
\newblock  (\bibinfo{year}{2016}).

\bibitem{Ma-16}
\bibinfo{author}{Ma, T.} \& \bibinfo{author}{Shvets, G.}
\newblock \bibinfo{title}{{All-Si valley-Hall photonic topological insulator}}.
\newblock \emph{\bibinfo{journal}{New Journal of Physics}}
  \textbf{\bibinfo{volume}{18}}, \bibinfo{pages}{025012}
\newblock  (\bibinfo{year}{2016}).

\bibitem{Rechtsman17}
\bibinfo{author}{Noh, J.} \emph{et~al.}
\newblock \bibinfo{title}{Topological protection of photonic mid-gap cavity
  modes}.
\newblock \emph{\bibinfo{journal}{arXiv: 1611.02373 [cond-mat.mes-hall]}}
\newblock  (\bibinfo{year}{2016}).

\bibitem{Regan16}
\bibinfo{title}{Direct imaging of isofrequency contours in photonic
  structures}.
\newblock \emph{\bibinfo{journal}{Science Advances}}
  \textbf{\bibinfo{volume}{2}}, \bibinfo{pages}{e1601591}
\newblock  (\bibinfo{year}{2016}).

\bibitem{Shen}
\bibinfo{author}{Shen, S.-Q.}
\newblock \emph{\bibinfo{title}{{Topological Insulators. Dirac Equation in
  Condensed Matters}}}
\newblock  (\bibinfo{publisher}{Springer}, \bibinfo{year}{2012}).

\end{thebibliography}

\end{document}